\documentclass[12pt]{article}
\newwrite\ffile\global\newcount\figno \global\figno=1

\def\writedef#1{}

\input epsf
\def\figin{\epsfcheck\figin}\def\figins{\epsfcheck\figins}
\def\epsfcheck{\ifx\epsfbox\UnDeFiNeD
\message{(NO epsf.tex, FIGURES WILL BE IGNORED)}
\gdef\figin##1{\vskip2in}\gdef\figins##1{\hskip.5in}
\else\message{(FIGURES WILL BE INCLUDED)}%
\gdef\figin##1{##1}\gdef\figins##1{##1}\fi}

\def\figinsert{}
\def\ifig#1#2#3{\xdef#1{fig.~\the\figno}
\writedef{#1\leftbracket fig.\noexpand~\the\figno}%
\figinsert\figin{\centerline{#3}}\medskip\centerline{\vbox{\baselineskip12pt
\advance\hsize by -1truein\center\footnotesize{  Fig.~\the\figno.} #2}}
\bigskip\endinsert\global\advance\figno by1}
\def\endinsert{}

\begin{document}
\baselineskip 18pt
\newcommand{\Tr}{\mbox{Tr\,}}
\newcommand{\beq}{\begin{equation}}
\newcommand{\eeq}{\end{equation}}
\newcommand{\bea}{\begin{eqnarray}}
\newcommand{\eea}[1]{\label{#1}\end{eqnarray}}
\renewcommand{\Re}{\mbox{Re}\,}
\renewcommand{\Im}{\mbox{Im}\,}
\begin{titlepage}

\begin{picture}(0,0)(0,0)
\put(350,0){SHEP-01-14}
\end{picture}
 
\begin{center}
\hfill
\vskip .4in
{\large\bf Secrets of the Metric In ${\cal N}$=4 and ${\cal N}$=2$^*$ Geometries}
\end{center}
\vskip .4in
\begin{center}
{\large James Babington, Nick Evans and  James Hockings}
\footnotetext{e-mail: jrb4@hep.phys.soton.ac.uk, 
n.evans@hep.phys.soton.ac.uk, jrh@hep.phys.soton.ac.uk }
\vskip .1in
{\em Department of Physics, Southampton University, Southampton,
S017 1BJ, UK}

\end{center}
\vskip .4in
\begin{center} {\bf ABSTRACT} \end{center}
\begin{quotation}
\noindent The metric of the gravity dual of a field theory should 
contain precisely the same information as  the field theory.
We discuss this connection in the ${\cal N}$=4 theory where a scalar 
vev may be introduced
at the level of 5d supergravity and the solutions lifted to 10d. We stress the
role of brane probing in finding the 
coordinates appropriate to the field theory. In these coordinates the
metric parametrizes the gauge invariant operators of the field theory
and either side of the duality is uniquely determined by the other.
We follow this same chain of computations for the 10d lift 
of the ${\cal N}$=2$^*$ geometry of 
Pilch and Warner. The brane probe of the metric reveals the 2d moduli 
space and the functional form of the gauge coupling. 
In the coordinates appropriate to the field theory the metric on moduli
space takes a very simple form and one can 
read off the gravity predictions for operators in the field theory.
Surprisingly there is logarithmic renormalization even in the far UV
where the field theory reverts to ${\cal N}$=4 super Yang-Mills. 
We extend the analysis of Buchel et al to find 
the D3 brane source distribution that generates the supergravity prediction for the 
gauge coupling 
for the whole class of solutions corresponding to different points on moduli 
space. This distribution does not account for the logarithmic behaviour in the
rest of the metric though. We discuss possible resolutions of the discrepancy. 

\end{quotation}
\vfill
\end{titlepage}
\eject
\noindent
\section{Introduction}

The $AdS$/CFT correspondence\cite{malda} has provided the 
first example of a fascinating duality 
between a strongly coupled gauge theory and a weakly coupled
gravity background. 
It has immediately been of interest to extend such dualities
to other gauge theories and gravity backgrounds to understand how generic
such a duality is. A number of techniques have been used to push forward
these explorations; finite temperature may be included by compactification
of the time direction \cite{w1,w2}, 
relevant deformations can be included by switching
on appropriate supergravity fields that act as sources in the ${\cal N}$=4 theory
\cite{gppz1}-\cite{bpp}, and
new D brane structures with different world volume theories and their near 
horizon geometries may be constructed \cite{larsen}-\cite{raja}. 
In this paper we want to make a deeper
investigation of the anatomy of some of these dualities. In principle two
theories which are dual should simply be reparametrizations of the same 
``solution''. Thus if we know the complete solution to some field theory
the corresponding gravity dual should be uniquely determined. Understanding
how this encoding occurs in some simple theories will hopefully lead to
new tools for constructing a wider class of dualities. In this paper we take
some steps in this direction.

In particular we will begin by revisiting the gravity 
duals of ${\cal N}$=4 super Yang
Mills on moduli space. The field theory is well understood and the 
gravity duals have been deduced both from D3 brane constructions 
\cite{larsen, brr}
and from
deformed 5d supergravity solutions \cite{freed2}
lifted to 10d solutions \cite{pw}. The latter solutions 
have been shown to match the former. Here we will approach this connection
from the field theory side. Starting with the 10d lift supergravity solution
we will show
that brane probing the solution provides a simple tool for
determining the unique 
coordinates it should be written in to make the gauge theory
correspondence manifest. The Dirac Born Infeld action of the probe provides a 
swift link between the gravity metric and the dimensionality of the field
theory moduli space and the functional dependence of the gauge coupling 
on that moduli space. In the special coordinates there is a manifest 
prescription for the encoding of the field theory operators in the metric. 
The 5d supergravity solutions only describe a subset of the possible moduli space
but using the prescription the full set of 10d supergravity solutions needed to
describe the full moduli space may be deduced. These metrics are indeed
solutions of the 10d supergravity equations of motion. 

To test whether the encoding prescription is generic we move these ideas
across to the gravity dual of the ${\cal N}$=2$^*$ theory (the N=4 theory with a mass
term that breaks supersymmetry to ${\cal N}$=2 in the IR) which has more interesting 
RG flow properties. Solutions produced by 
including relevant deformations in the 5d supergravity theory \cite{gub2, bs2, ep}
have been lifted 
to 10d by Pilch and Warner \cite{pw} and also in \cite{bs2}. 
The conection to the gauge theory of this set 
of solutions is far
from apparent after the lift. The use of a brane probe to uncover
the links was made in \cite{ejp} and \cite{bpp}. 
The metric indeed describes the
expected  2d moduli space of the field theory. The gauge 
coupling function on the moduli space is also revealed and, when 
the solution is placed in appropriately ${\cal N}$=2 coordinates \cite{bpp},
matches to field 
theory expectations. The set of solutions describe different points on moduli
space with one corresponding to a singular point on moduli space where in the 
IR the gauge coupling diverges. This solution is of interest because it
provides an example of the enhan\c con mechanism \cite{pjp}
(there are points in the space
where a probe's tension falls to zero). We write the metric on moduli space
in the
coordinates applicable to the field theory where it takes the form of 
a single function as in the ${\cal N}$=4 metrics multiplied by the gauge coupling 
function. It is natural to interpret the outstanding function according 
to the same prescription as in the ${\cal N}$=4 solution and read off field theory
operators. In the field theory the gauge coupling encodes the only 
renormalization group (RG) flow \cite{nonren}
whilst the supergravity solution appears to describe additional renormalization
of the scalar operators. In addition in the far UV the solution does not
return to the ${\cal N}$=4 form but contains logarithmic renormalization. 
To highlight the discrepancy we follow the prescription in \cite{bpp}
for deducing the D3 brane distribution from the expected field theory gauge
coupling, as a function of position on moduli space, and the supergravity form for
the coupling. We thus deduce the distribution for all the 5d supergravity lifts
and can then calculate the expected scalar operators which again do not 
match with the function in the metric. Presumably there is some 
discrepancy in the prescription in this more complicated theory we have not
yet discovered, nevertheless, we believe the approach to be valuable and
the discrepancies an important starting point for future progress.

\section{The Gravity Dual of ${\cal N}$=4 on Moduli Space }

We begin by studying gravity solutions describing ${\cal N}$=4 
super Yang-Mills (SYM) theory  on moduli space
resulting from 5d supergravity \cite{freed2}. 
We wish to study the gauge theory in the presence 
of a non-zero vacuum expectation value (vev) 
for the scalar operator $tr \phi^2$. Since there are
6 real scalars this operator is a 6x6 matrix with the symmetric traceless
entries transforming as the 20 of the global $SU(4)_R$ symmetry of the theory.
In the 5d truncation of IIB supergravity on $AdS_5 \times S^5$ \cite{warn, grw2}
the lightest state
is a scalar, $\alpha$, in the 20 that acts as the 
source for $tr \phi^2$ in the $AdS$/CFT correspondence \cite{gkp, w1}. 
One may look
for solutions of the 5d supergravity equations of motion with non-zero $\alpha$
and interpret them as gravity duals of the ${\cal N}$=4 theory with a scalar vev
switched on. In fact, considerable work is needed to arrive at the equations
of motion since the scalars live in the coset $E_6/USp(8)$, the subtleties 
of which are discussed in \cite{freed2}. 
We shall present the final results only here.

As an example let us consider the case of switching on $tr \phi^2 = {\rm diag}
(1,1,1,1-2,-2)$. The appropriate supergravity scalar has been identified in 
\cite{freed2}. In
the supergravity theory the metric is dynamical and the scalar vev cannot be 
considered in isolation. We parametrize the metric as
\begin{equation}
ds^2=e^{2A(r)} dx_{//}^2-dr^2
\end{equation}
where $x_{//}$ describe four dimensional Minkowski 
space slices through the deformed $AdS$ space, 
$r$ is the radial direction, and in the $AdS$ limit $A(r) = r/L$ with $L$ the 
radius of the $AdS$ space.
The resulting supersymmetric equations of motion 
(for which the fermionic shifts vanish) 
are first order(where $\rho=e^{\alpha}$),
\begin{equation}
\frac{\partial\rho}{\partial r}=
\frac{1}{3L}\left(\frac{1}{\rho}-\rho^5\right), 
\hspace{1cm} \frac{\partial 
A}{\partial r}=\frac{2}{3L}\left(\frac{1}{\rho^2}+\frac{\rho^4}{2}\right).
\end{equation}
These equations may be solved in the $\rho-A$ plane since
\begin{equation}
{\partial \rho \over \partial A} = {1 \over 2} \left( {\rho - \rho^7 \over
1 + \frac{\rho^6}{2}} \right) 
\end{equation}
with solution
\begin{equation} \label{Atorho}
e^{2A} = \frac{l^2}{L^2} {\rho^4 \over \rho^6-1}
\end{equation}
with $l^2/L^2$ a constant of integration. At this level the connection to
the dual gauge theory is somewhat opaque.

Remarkably the solution has been lifted 
back to a $D=10$ solution \cite{freed2, pw}
which takes the form

\begin{equation} \label{4met}
ds^2=\frac{X^{1/2}}{\rho}e^{2A(r)} dx_{//}^2 - 
\frac{X^{1/2}}{\rho}
\left(dr^2+\frac{L^2}{\rho^2}\left[d\theta^2+\frac{\sin^2\theta
}{X
}d\phi^2+\frac{\rho^6\cos^2\theta}{X}d\Omega^2_3\right] \right),
\end{equation}
where $d\Omega^2_3$ is the metric on a 3-sphere and
\begin{equation}
X  \equiv  \cos^2\theta+\rho^6\sin^2\theta
\end{equation}
For consistency there must also be a non-zero $C_4$ potential of the form
\begin{equation} \label{4C4}
C_4 = \frac{e^{4A}X}{g_s  \rho^2} dx^0 \wedge dx^1 \wedge dx^2 \wedge dx^3
\end{equation}

Again any relation to a dual theory is well hidden. In fact in \cite{freed2}
this metric was determined to be equivalent to the near horizon limit
of a multi-centre solution around a D3 brane distribution.
We wish to make the need for the transformation to these coordinates
clear within the context of the duality. There are many coordinate 
redefinitions one could make and only a single set of coordinates can
manifestly display the field theory duality. A tool
is needed to find these coordinates and the appropriate choice for that tool,
as we will see, is brane probing.  

Brane probing \cite{cliffnotes}
is most transparent in the orginal D3 brane construction
for the $AdS$/CFT correspondence. 
Here there is a stack of N D3 branes at the origin 
with the ${\cal N}$=4 SYM as their world volume theory and $AdS_5\times S^5$  as their
near horizon geometry. If we imagine moving a single D3 brane from
the stack and moving it in the space then, to first order, it will not effect
the background metric. From the world volume field theory point of view, by
separating a D3 brane we have introduced an adjoint scalar 
vev breaking SU(N)$\rightarrow$
U(1)$\times$SU(N-1). The magic of D-branes is that the scalar 
fields' vevs in the field theory are precisely identified 
with the position of the D3 brane in the surrounding spacetime. 
This is expressed by the Dirac Born Infeld (DBI) action for a D3 brane,
\begin{equation} \label{BI}
S_{probe}=-\tau_3\int_{\mathcal{M}_4}d^4x
\det[G^{(E)}_{ab} + 2 \pi \alpha' e^{- \Phi/2} F_{ab}]^{1/2} 
+ \mu_3\int_{\mathcal{M}_4} C_4,
\end{equation}
where $G_{ab}$ is the pull back of the spacetime metric, $F^{ab}$ the gauge
field on the probe's surface, $\Phi$ the dilaton (which is
a constant in this solution) and $\tau_3 = \mu_3/g_s$.
Thus the DBI action allows us to translate the background metric to 
a potential for the scalar fields in the field theory. It is easy 
to identify the dimension of the field theory moduli space implied by
the metric from where the DBI potential vanishes.
In addition since the U(1) theory lives on the probe's surface
and is a non-interacting theory (photons do not self interact and
there is  only adjoint matter which for a U(1) is chargeless), its coupling
is that of the SU(N) theory at the scale of the breaking vev. The probe also 
therefore lets us determine the functional form of the coupling on moduli
space.

We proceed to brane probe the 10d metric above by substituting 
(\ref{4met})-(\ref{4C4}) in (\ref{BI}).
Allowing the brane to move slowly and concentrating on the scalar sector,
we find the DBI action corresponds to the field theory,
\begin{equation} \label{notflat}
S=-{\mu_3  \over 2 g_s}
\int_{\mathcal{M}_4}d^4x \left[
\frac{Xe^{2A}}{\rho^2}\dot{r}^2+\frac{L^2 e^{2A}}{\rho^4}
(X \dot{\theta}^2+\sin^2\theta ~ \dot{\phi}^2
+\rho^6\cos^2\theta ~ \dot{\Omega}^2
_
3) \right].
\end{equation}
The immediate result is that we see there is no potential against motion
of the probe in the full 6 dimensional transverse space corresponding in
the field theory to the scalars having a 6d moduli space. This matches with 
our expectations for the ${\cal N}$=4 SYM theory where the six scalars have a potential
of the form $tr[\phi,\phi]^2$ and so, taking commuting vevs, the six scalars
may take arbitrary values. 

The kinetic terms should be interpreted as the kinetic terms of the field
theory scalars which in the ${\cal N}$=4 theory are given by $(1/8 \pi) Im (
\tau \Phi^\dagger \Phi)|_D$
(in ${\cal N}$=1 notation). The coefficient of the kinetic terms are therefore the
gauge coupling which is known to be conformal in the ${\cal N}$=4 theory. We should
expect the metric that the probe sees on moduli space to be flat which it
manifestly isn't in (\ref{notflat}). 
This is our hint as to the coordinate change we should
make in order to pass to those coordinates 
where the duality is manifest. Forcing this relation we find a change of coordinates that makes the probe metric flat 
\begin{equation}
(r,\theta)\rightarrow (u,\alpha)
\end{equation}
such that
\begin{equation} \label{coord}
u^2\cos^2\alpha=L^2e^{2A}\rho^2\cos^2\theta, \hspace{1cm}
u^2\sin^2\alpha=L^2\frac{e^{2A}}{\rho^4}\sin^2\theta.
\end{equation}
A small calculation shows that the metric in these coordinates takes the form
\begin{equation}
S  = -{\mu_3 \over 2 g_s}
\int_{\mathcal{M}_4} d^4x  \left[ \dot{u}^2+u^2(\dot{\alpha}^2+ 
\sin^2\alpha  ~ \dot{u}^2 + \cos^2\alpha  ~  \dot{\Omega}^2_3) 
\right].
\end{equation}

This is a unique choice of coordinates and if the duality is to be 
manifest it must be in these coordinates where the coupling is seen
to have the correct conformal property. It is therefore interesting to write
the full metric in these coordinates

\begin{equation} \label{prenice}
ds^2= \left(\frac{\rho^2}{Xe^{4A}}\right)^{-1/2}
dx_{//}^2   +   \left( \frac{ \rho^2
}{X
e^{4A}} \right)^{1/2}   \sum_{i=1}^{6}(du_i)^2 \label{met1}
\end{equation}
This is of the familiar form,
\begin{equation} \label{nice}
ds^2=H^{-1/2} dx_{//}^2+H^{1/2}\sum_{i=1}^{6}du_i^2, \hspace{1cm}
C_4 = {1 \over H g_s} dx^0 \wedge dx^1 \wedge dx^2 \wedge dx^3
\end{equation}

From the coordinate transformations (\ref{coord}) and using (\ref{Atorho}), 
we can obtain an explicit expression 
for $\rho$ in terms of $(u,\alpha)$
\begin{equation} \label{rhoofu}
{u^2 \over l^2} \sin^2 \alpha ~\rho^{12} + \left({u^2 \over l^2} \cos^2 \alpha
- {u^2 \over l^2} \sin^2 \alpha -1\right) \rho^6 - 
{u^2 \over l^2} \cos^2 \alpha = 0
\end{equation}

In \cite{freed2} it was shown that in these coordinates 
$H(u)$ can be written as a multi-centre solution with a D3 density, $\sigma$,
\begin{equation} \label{n4h}
H(u)=\int d^6x ~ \sigma(x) ~ \frac{L^4}{|\vec{u}-\vec{x}|^4}
\end{equation}
In this case the density is a 2 dimensional disk of uniform density
in the $\theta = \pi/2$ plane
\begin{equation}
\sigma(x) = \frac{1}{\pi l^2}\theta(l^2-x^2) 
\end{equation}

We wish to  make the connection to the field theory and instead 
consider the large $u$ limit of (\ref{rhoofu}) from which we obtain
\begin{equation}
\rho^6=1+\frac{l^2}{u^2}+\left(\frac{l^2}{u^2}\right)^2(1-\sin^2\alpha)+\left(
\frac{l^2}{u^2}\right)^3(1-3\sin^2\alpha+2\sin^4\alpha)  + 
\mathcal{O}(\frac{l^8}{u^8}).
\end{equation}
and hence from (\ref{prenice})

\begin{equation}\label{finH}
H(u)=\frac{L^4}{u^4}\left(1+\frac{l^2}{u^2}
(3\sin^2\alpha-1)+\frac{l^4}{u^4}(1-8\sin^2\alpha+10\sin^4\alpha)\right)+ 
\mathcal{O}(\frac{L^4l^6}
{u^{10}}). 
\end{equation}

In this form it is possible for us to identify field theory operators 
\cite{larsen}.
The radial coordinate $u$ has the scaling dimension of mass \cite{w1} 
so in each term in
the expansion we can assign a scaling dimension to the coefficient. 
Further each term in the expansion is associated
with a unique spherical harmonic\footnote{The spherical harmonics may 
be found by writing the 6 dimensional representation as a unit vector
in the transverse space 
and then finding the symmetric traceless products $6 \times 6 = 20 +..$,
$6 \times 6 \times 6 \times 6= 50 + ...$,  etc}; 
the angular function in the $1/u^6$ term is
the spherical harmonic in the 20 of SU(4)$_R$, that in the $1/u^8$ term
the harmonic in the 50 and so forth. Note that by using the orthonormality
of the spherical harmonics it is easy to show that each harmonic occurs only 
in a single term in the expansion. We can therefore identify the $n$th
coefficient as having the dimension and symmetry properties of the field
theory operator $tr \phi^n$ and further that the operator is not 
renormalized since there is no further function of $u$ associated with
the operator. Thus these solutions suggest the general
form 
  
\begin{equation}
H(u)=\frac{L^4}{u^4}( 1 + \sum_n \frac{tr \phi^n}{u^n} Y_n). 
\end{equation}
where $Y_n$ is the spherical harmonic obtained from the product of $n$ 
6 dimensional reps. 

It is worth noting that at the level of the 5d supergravity theory we introduced
only a vev for the dimension 2 operator $tr \phi^2$ yet after the
lift to 10d the solution was forced to possess vevs for higher 
dimension operators. If we returned to 5d the truncation would again
remove these operators. 
The 5d supergravity metric gives specific relations between the operators as
is explicit in (\ref{finH}) 
whilst in the field theory they are expected to be arbitrary
reflecting the 6 dimensional moduli space.  One may therefore try substituting 
the expansion with arbitrary coefficients into the supergravity 
field equations 
and they indeed turn out to be solutions \cite{larsen}. 
Of course in this context this is no surprise because it is already 
known that the multi-centre solutions are solutions of the field equations
for arbitrary D3 brane distributions. However, it is encouraging in this 
simplest case that one can deduce a full gravity description of the field
theory from the 5d supergravity solutions. Further it is appealing that the 
metric is indeed seen to be a rewriting of the field theory solutions
and it is of interest to see how this generalizes in theories with more
complicated RG flow. In the next section we will study aspects of this 
generalization for the ${\cal N}$=2$^*$ theory.

Before moving on though we wish to note the power of the brane probing
technique since it in fact is capable of deriving the above solutions 
on its own. In the ${\cal N}$=4 case if we wished to write down a metric dual
to a point on moduli space we might begin by  writing down an arbitrary 
10d metric. If we then require the 6 dimensional moduli space and conformal
coupling after a brane probe the metric is forced to take the form 
in (\ref{nice}). 
The supergravity field equations 
with this ansatz reduce to the transverse flat space 
Laplacian in 6 dimensions \cite{petersen},
\begin{equation}
\triangle_6H(u)= 0
\end{equation}
Which produces the multi-centre solutions. We see again that 
when we know sufficient information
about the field theory the supergravity dual is uniquely determined.

\section{The ${\cal N}$=2$^*$ Geometry}

We have seen that in the ${\cal N}$=4 duality there is a simple mapping between
the field theory operators and the form of the metric. It would be interesting
to understand how this mapping occurs in a more complicated theory with
non-trivial renormalization group flow. The theory we choose to investigate
in this light is the ${\cal N}$=2$^*$ theory where a mass term is introduced into 
the ${\cal N}$=4 theory that leaves an ${\cal N}$=2 supersymmetric theory in the IR. The 5d
supergravity theory with the appropriate supergravity field deformations switched on
was studied in \cite{gub2, bs2, ep}. 
Two supergravity scalars are needed, one describing the mass
term and the other the possible vev for the remaining two real scalar fields.
Although some connections were made between the field theory 
and these solutions the duality remained fairly opaque at the 5d level. 
A lift of this solution to 10d supergravity has again been provided 
\cite{pw,bs2} and the
summary of the solution is
\begin{equation}
ds^2 = \Omega^2 ( e^{2  A} dx_{//}^2 + dr^2) + {L^2 \Omega^2 \over \rho^2}
( {d \theta^2 \over c} 
+ \rho^6 \cos^2 \theta  ( {\sigma_1^2 \over  c X_2} + 
{\sigma_1^2 + \sigma_2^2 \over X_1})+ {\sin^2 \theta
\over X_2} d \phi^2 )
\end{equation}
where
\begin{equation}
\Omega^2 = ( c   X_1 X_2 )^{1/4} / \rho, \hspace{1cm} c = \cosh 2 m
\end{equation}
\begin{equation}
X_1 = \cos^2 \theta + \rho^6 c \sin^2 \theta, \hspace{1cm}
X_2 = c \cos^2 \theta + \rho^6  \sin^2 \theta
\end{equation}
\begin{equation}
C_4 = { e^{4A} X_1 \over g_s \rho^2} dx^0 \wedge dx^1 \wedge dx^2 \wedge 
dx^3 
\end{equation}
The dilaton is non-trivial too. We write a complex scalar
$ \lambda = C_0 + i e^{- \Phi}$ and

\beq
\lambda = i \left( { 1 - B \over 1 + B} \right), \hspace{1cm}
B = \left( { b^{1/4} - b^{-1/4} \over b^{1/4} + b^{-1/4} } \right),
\hspace{1cm} b = \cosh (2m) {X_1 \over X_2}
\eeq

The fields $m$, $A$ and  $\rho = e^\alpha$ are the supergravity fields given by
the 5d supergravity equations of motion 

\begin{eqnarray}
{ \partial \alpha \over \partial r} & = & {1 \over 3L} \left( {1 \over \rho^2} 
- \rho^4 \cosh(2 m) 
\right)\\
 { \partial A \over \partial r} & = & {2 \over 3L} 
\left( {1 \over \rho^2} + {1 \over 2} \rho^4\cosh(2 m)
\right)\\
 { \partial m \over \partial r} & = & - {1 \over 2L} 
\rho^4 \sinh ( 2 m)
\end{eqnarray}
which have solutions

\beq 
e^A = k { \rho^2 \over \sinh ( 2 m) }
\eeq

\beq \label{rho}
\rho^6 = \cosh ( 2 m ) + \sinh^2 (2 m) \left( \gamma + \log \left[ 
{\sinh m \over \cosh m } \right] \right)
\eeq

The solution also has non-zero 2-forms \cite{pw} but they are zero in the 
$\theta = \pi/2$ plane which we will analyze below.

Again brane probing has provided the first deep insight into the duality
with the field theory. In \cite{ejp} and \cite{bpp}
it was observed that after substituting 
the above 10d solution into the DBI action the potential vanishes 
in the $\theta = \pi/2$ plane. The moduli space for brane motion therefore
matches the expected 2d moduli space of the ${\cal N}$=2$^*$ field theory which 
has two massless real scalars. From now on we will restrict our attention to
this plane. Placing a brane probe off the moduli space corresponds in the 
field theory to giving a vev to a massive scalar which is neither a vacuum
of the theory nor supersymmetric. We know of no field theory results in
the presence of such vevs so there are no checks of the duality we can make.

On the moduli space a brane probe reveals the U(1) field theory
\beq
{\cal L} = {1 \over 2} \left(\rho^4 \cosh (2m) e^{2A} \dot{r}^2
+ {L^2 \rho^4 \cosh (2m) e^{2A} \over \rho^8} \dot{\phi}^2 \right)
+ {1 \over 4} \tau_3 ( 2 \pi \alpha^\prime)^2 e^{-\Phi}
F^{\mu \nu} F_{\mu \nu} 
\eeq

In these coordinates the connection to the ${\cal N}$=2$^*$ theory is hidden 
but we can now find coordinates where the duality is manifest.
The two scalar fields should have the same kinetic term with a common
coefficient given by the gauge theory's running coupling, $1/g^2_{YM}(r)$.
The first of these can be achieved by the change of coordinates

\beq \label{vcoord}
v = L \sqrt{ \cosh 2m+1 \over \cosh 2m-1}
\eeq
such that

\beq
{\partial v \over \partial r} = {\rho^4 \over L} v 
\eeq
and we have

\beq \label{vlag}
 {\cal L} = { 1 \over 2} {k^2 L^2  \cosh 2m \over \sinh^2 2m ~~ v^2} (\dot{v}^2
+ v^2 \dot{\phi}^2)
\eeq

The solutions depend on two constants $k$ and $\gamma$ which correspond
to the mass term and the scalar vev \cite{bpp} respectively.
It is interesting to  discuss the anatomy of these solutions at fixed $k$
as a function of $\gamma$ in the $v$ coordinates. 
As in previous work \cite{gub2, ep, ejp, bpp} we only consider
$\gamma \leq 0$ since we can offer no physical interpretation of
positive $\gamma$. 
Although, as we will see, $v-\phi$ 
are not the physical coordinates for the duality
they have the benefit of an SO(2) symmetry in $\phi$ as can be
seen from (\ref{vlag}). 
The solutions with different choice of the
parameter $\gamma$ differ in the radial position at which the metric 
has divergences as a result of $\rho \rightarrow 0$. From (\ref{rho}) and 
(\ref{vcoord}) one
may express $\gamma$ in terms of this radius $l$ as

\beq
\gamma = -{l^2 \over 4 L^2} + {L^2 \over 4 l^2} + \ln l/L
\eeq

We expect the divergence in the metric to be associated with the presence
of a disc D3 brane source and hence solutions with larger negative $\gamma$ 
correspond to larger vevs in the field theory. When $\gamma=0$ the
spacetime is good down to a radius $v=l=L$ where $\cosh 2 m \rightarrow \infty$
and hence the coeffcient of the scalar kinetic term falls to zero. 
This is the enhan\c con locus where the probe's tension falls to zero 
(or in the field theory the coupling diverges)
and according to lore \cite{pjp} 
we must excise the solution within. Only for this 
metric can the enhan\c con point be reached since the other, $\gamma < 0$,
solutions have 
$\rho \rightarrow 0$ at a larger radius where the scalar kinetic terms 
coefficient is still regular.

%
%
%

As pointed out in \cite{bpp} 
we can not yet formally make the connection to the 
gauge coupling because the U(1) theory is not in an ${\cal N}$=2 form since
the coeffcient of the $F_{\mu\nu}^2$ term is given by

\beq
e^{- \Phi} = {c \over g_s |\cos \phi +ic \sin \phi|^2  }
\eeq

To obtain an ${\cal N}$=2 form 
we must make a conformal transformation in the 
$v-\phi$ plane to  equate the coefficients of the scalar and gauge field
kinetic terms. The transformation is \cite{bpp}
\beq \label{ycoord}
Y = {k L \over 2} ({V \over L} + {L \over V})
\eeq
where $V= ve^{i \phi},Y = y e^{i \eta}$ are 
complex parameters on the 2d space. The low energy theory is then of 
the desired form with
\beq \label{us}
{\cal L} = { 1 \over  g^2_{YM}(Y)} |\dot{Y}|^2  + {\rm Im} \left(
\tau (F^{\mu \nu} 
F_{\mu \nu} + iF^{\mu \nu}\tilde{F}_{\mu \nu})  \right)
  \eeq
with  $4 \pi/g^2_{YM}(Y)  = Im \tau$ where
\beq\label{supergravitytau}
\tau = {i \over g_s} \sqrt{ Y^2 \over Y^2 - k^2 L^2}
\eeq

In these coordinates the background takes the form

\beq \label{N2metric}
ds^2 = {1 \over g_{YM}} \left(
H^{-1/2} dx_{//}^2 + H^{1/2}  dy^2 \right), \hspace{0.5cm}
C_4 = {g^2_{YM} \over Hg_s} dx^0 \wedge dx^1 \wedge dx^2 \wedge dx^3, \hspace{0.5cm} 
\tau_3 (2 \pi \alpha^\prime)^2 e^{-\Phi} = {1 \over g^2_{YM}}
\eeq
with

\beq \label{n2h}
g^2_{YM} H = {\sinh^4 2m\over k^4   \rho^{12} \cosh 2m}
\eeq

All other fields are zero in the $\theta = \pi/2$ plane.
In fact the brane probe does not uniquely fix the form of $H$ since 
it can be rescaled by an arbitrary power of the Yang Mills coupling 
and still return the same probe theory. 
Since the coupling in (\ref{supergravitytau})
does not contain logarithms such a rescaling will not resolve the 
discrepancies discussed below. 

We claim to have identifed the unique coordinates in which in the 
$\theta = \pi/2$ plane a brane probe correctly matches the expected form for
an ${\cal N}$=2 supersymmetric theory. In these physical coordinates we would
expect the remainder of the metric to be a parametrization of field 
theory operators. To see the predictions for these operators we can 
expand the $H$ function  at large radius in these coordinates.

We note that the final transformation in (\ref{ycoord}) is rather strange 
since the circle $v=L$ is mapped to the real line of length $2kL$ and 
everything interior is mapped 
to exterior points to the line in $Y$ space. Thus the
$V$ coordinates are a double cover of the $Y$ space. In the 
$v$ coordinates one can not take a probe through the enhan\c con so one should
exclude the region $v < 1$. 

At large $y$ the $v$ coordinate, from (\ref{ycoord}), is given by 

\beq
v = {2 y \over k} - {k \cos 2 \eta \over 2y} + 
{k^3 \over 32 y^3} (1-5 \cos 4 \eta) +.... 
\eeq

Thus at large $y$ we find, using (\ref{rho}) (\ref{ycoord}) and (\ref{n2h})

\begin{eqnarray} \label{supergravityH}
H & = & {L^4 k^4 \over 16 y^4}  \hspace{0.3cm}
+ \hspace{0.3cm} {L^6 k^6 \over 64 y^6} ( -2 + 2 
{l^2 \over L^2}-  {2 L^2\over l^2} + 8 \ln (y/l)
+ 6 \cos(2 \eta) ) \nonumber \\
&&
+ {L^8 k^8 \over  2^8 y^8} \left[ 3 ( 1 -{l^2 \over  L^2} 
+ {L^2 \over  l^2} + 4 \ln (y/L) -2  \cos 2 \eta)^2 
+ 2 \cos 2 \eta ( -2 + 2 
{l^2 \over L^2}-  {2 L^2\over l^2} + 8 \ln (y/l))\nonumber \right.\\
&& \left. \right. \hspace{2cm} \left.
+ ( 3 +2 {l^2 \over  L^2} 
- 2{L^2 \over  l^2} - 8 \ln (y/L) - 8 \cos 2 \eta + 14 \cos 4 \eta 
) \right] + ...
\end{eqnarray}

Finally we have arrived at the form for the metric we're interested
in. The metric on moduli space, when written in the physical coordinates
that explicitly display ${\cal N}$=2 supersymmetry 
in the brane probe, has two functions in it.
One is the gauge coupling of the theory and the other, $H$, remains to be
interpreted. We can read off the symmetry properties of operators from
$H$ using the same prescription as for the ${\cal N}$=4 solution; 
every factor of $y$ carries mass dimension 1 and the $\eta$ dependence
can be interpreted as SO(2) harmonics $\cos n \eta$ with charge $n$. Thus
one would naturally like to interpret the coefficient of $\cos n \eta$,
which has U(1) charge $n$,
as the operator $tr \phi^n$ 
(with $\phi$ the massless, two component, complex scalar field) 
and would expect it to be 
associated with a factor of $y^{(n+4)}$. The charge zero coefficients 
would correspond to $tr |\phi|^{n}$ again associated with a factor
of $y^{n+4}$. There are also mixed operators of the form of a product
of these two operator types as can be seen from the presence of a 
$\cos 2 \eta$ term at order $1/y^8$.
The presence of logarithms though undermines this interpretation. In the $l
\rightarrow \infty$ limit one would expect the ${\cal N}$=2$^*$ theory to be on 
the edge of its moduli space and return to looking like the ${\cal N}$=4   
metric. In fact at large $l$ the leading terms in $l$ do indeed take the form
in (\ref{finH})
but we can not neglect the $\log y$ terms in this limit which are absent
from the ${\cal N}$=4 theory. There appears therefore to be UV logarithmic 
renormalization. Given that there is logarithmic renormalization we can 
not rule out power like renormalization either which would further 
confuse the interpretation.

We will make this discrepancy more manifest in the next section where we 
deduce the D3 brane distributions from the form of the gauge coupling 
and show that it does not predict the above form for the field theory
operators. In the discussion we will suggest a few possible resolutions
of the discrepancy.

\subsection{D3 Distributions} 

To highlight the discrepancy between field theory expectations and 
the $H$ function found in the ${\cal N}$=2$^*$ metric we will determine the
D3 brane distribution function for spacetimes with different $\gamma$
assuming the standard one loop renormalized expression for the 
prepotential governing the IR of the theory.  
The field theory is reviewed in \cite{bpp} and 
the authors followed this logic for the special 
case $\gamma = 0$, where in $Y$ space the D3 branes are distributed on a 
line.
We extend the analysis to all $\gamma$.
The prepotential for the ${\cal N}$=2$^*$ theory is expected to be 
\beq 
{\cal F} = { i \over 8 \pi} \left[ \sum_{i\neq j} (a_i - a_j)^2 \ln \left(
{(a_i-a_j)^2 \over \mu^2 }\right) -  \sum_{i\neq j}(a_i - a_j + m)^2 \ln \left(
{(a_i-a_j+m)^2 \over \mu^2 }\right)\right]
\eeq
where $a_i$ are the scalar vev eigenvalues and $\mu$ an RG scale.
In the supergravity description of the 
N=2$^*$ theory we expect the scalar vevs to be large with respect to 
the mass term and hence \cite{bpp}
\beq \label{tau}
\tau(Y) = {i \over g_s} + {i \over 2 \pi} \int \sigma {m^2 \over (Y-a)^2} d^2a
\eeq
where $a$ is a complex 2d integral in $Y$ space and $\sigma$ the density of
vevs/D3 branes. To match with the supergravity we make the identification 
$m^2 = k^2 \pi / L^2$ \cite{bpp}.
Using this ansatz we can determine the distributions that reproduce
the supergravity solution expression for $\tau$. In fact this is all but 
impossible in $Y$ space since there is no spherical symmetry 
but we know that in $V$ the distributions are circular out to $l$
and cut off inside at $v=L$. 

Remarkably, a simple form for the denisity, $\sigma$, for each of the 
solutions, labelled by $\gamma$ or equivalently $l$, can then be found  
by rewriting (\ref{tau}) in $V$ space using (\ref{ycoord})
and using

\beq
\sigma_v v dv d\phi  = \sigma_y y dy d\eta
\eeq

Expanding the resulting expression as a power series at large $y$ and inserting
an expansion in powers of $1/v$ 
for $\sigma_v$ one can show to all orders in the expansion
that

\beq
\sigma_v = {1 \over  
\pi (l^2 - L^4/l^2)} (1 + L^4/v^4 - 2 L^2 \cos(2 \phi)/v^2)
\eeq
reproduces the supergravity expression (\ref{supergravitytau}). Note that this result agrees
with that of \cite{bpp} for $\gamma = 0$, $l=L$; integrating with a 
measure $v dv$ from $v=L$ to $l$
and then taking the $l \rightarrow L$ limit we obtain an expression
for the number of D3 branes of the form

\beq
N_{D3} = {1 \over \pi} \int^\pi_0 (1- \cos 2 \theta) d\theta
\eeq
Changing variables to $y= k L \cos \theta$ this reproduces the
line density in \cite{bpp}

\beq
\sigma_y = { 2 \over m^2 } \sqrt{k^2 - y^2} 
\eeq

Having identified the  density we can then predict the expected 
scalar operators. Since the only renormlization in the ${\cal N}$=2$^*$ theory
is that of $\tau$ \cite{nonren} we would expect the ${\cal N}$=4 expression for the 
metric quantity $H$
when evaluated in the $\theta=\pi/2$ plane
to display the full set of operators. Thus using 
(\ref{n4h}) (with $y$ rescaled to $2y/k$),
performing the integration after a change of variables
to $V$ space using
\beq
y \cos\eta = {kL \over 2} (v + 1/v) \cos \phi, \hspace{1cm}
y \sin \eta = {kL \over 2} (1/v-v)\sin \phi
\eeq
\beq
y^2 = {k^2 L^2\over 4} ( v^2 + 1/v^2 + 2 \cos \phi - 2 \sin \phi)
\eeq
and further expanding at large $y$ and evaluating the expression 
in the $\theta = \pi/2$ plane we obtain a prediction for $H$

\begin{eqnarray}
H & =& {L^4 k^4 \over 16 y^4} \hspace{0.3cm} + \hspace{0.3cm}
{L^6 k^6 \over 64 y^6} (  {2l^2 \over L^2} + { 2L^2 \over l^2} 
+ 6 \cos(2 \eta) ) \nonumber \\
&&+ {k^8 L^8 \over 2^8 y^8} \left( 3 ( {l^4 \over L^4}
+ 4 + {L^4 \over l^4}) + 16 {l^2 \over L^4} (1 + {L^4 \over l^4}) \cos 2 \eta
+ 20 \cos 4 \eta\right)
\end{eqnarray}

This expression does not match that in (\ref{supergravityH})
highlighting the apparent discrepancy in the interpretation of the 
coefficients as the scalar operators. There appears to 
be extra logarithmic and power renormalization in the supergravity theory
that this simple field theory analysis has not explained.

\section{Discussion}
 
Our philosophy has been to try to understand how supergravity solutions dual
to field theories encode the field theory operators and their running,
concentrating on gravity duals obtained by deforming the 5d supergravity
$AdS$/CFT correspondence.
One would expect to be able to interpret all elements of the supergravity solution
in this light. In the ${\cal N}$=4 theory this is indeed the case with the metric
being a simple encoding of the scalar operator vevs. To extend this 
understanding to theories with more interesting renormlization properties
requires undestanding the complicated procedure of including relevant
deformations in 5d supergravity, lifting the solutions to 10d and then finding
the physical coordinates appropriate to the duality. We have first persued this
chain of analysis 
for the ${\cal N}$=4 theory on moduli space with the scalar vevs introduced
as deformations. Brane probing has proved to be the vital tool for identifying
the physical coordinates; the probe allows us to identify the moduli space
and the functional form of the gauge coupling on that moduli space. Matching
the gauge coupling to field theory expectations in the ${\cal N}$=4 theory provides
the physical coordinates in which the encoding of the field theory operators
are manifest. This prescription then allows the class of solutions
to be extended to describe the full moduli space of the theory. 

Armed with this tool we have applied it to the ${\cal N}$=2$^*$ gravity dual. 
Brane probing the solution
reveals the 2d moduli space and, identifying the unique coordinates 
in which the U(1) theory on the probe takes an ${\cal N}$=2 form, 
the gauge coupling on that moduli space. These should be the physical 
coordinates in which the duality to the field theory is manifest in the
rest of the metric. The metric indeed takes a form on 
the moduli space analagous
to the metric on moduli space in the ${\cal N}$=4 theory except that the 
running of the gauge coupling is also encoded. 
There is one other function in the metric from which we can read off operators
by their scaling dimension and their symmetry properties. In the field theory
we expect the gauge coupling to be the only renormalized quantity and the 
operators $tr \phi^n$ and $tr |\phi|^n$ 
to emerge as in the ${\cal N}$=4 case. In fact we find further
renormalization including UV logarithmic renormalization. 

The appearance of this extra renormalization is frustrating because it
stops us from completely understanding the prescription for creating
a gravity dual to a field theory even in the next simplist case to the ${\cal N}$=4
theory. The form of the metric on moduli space in (\ref{N2metric}) 
is highly suggestive
that the prescription is to encode the running coupling
as shown and then parametrizes the scalar vev operators
in the field theory through $H$. 
It may be that the discrepancies we have seen are 
complications brought in by the 5d supergravity approach to
constructing the dualities. One possibility is that we have not only
introduced a mass term into the field theory. In the ${\cal N}$=4 theory when one
attempts to introduce a dimension 2 operator at the level of 5d supergravity,
after the lift to 10d, a whole host of higher dimension operators are
found to be present to make the solution consistent (as can be seen 
in (\ref{finH})). Something similar
may be happening here and the ${\cal N}$=2$^*$ solution is encoding both the field 
theory scalar vevs and this unknown tower of deformations. 

An alternative 
possibility is that the 5d supergravity solution was created in the coordinates $V$
which are a double cover of the physical coordinates $Y$. We have 
excised the solution interior to $v=L$ but possibly there is additional
interior structure which in the $Y$ coordinates is projected to large
$y$. Possibly in the physical coordinates there are D3 branes through out
the whole space! 

In spite of these obstacles we believe the philosophy of the paper,
and its support from the ${\cal N}$=4 solutions and the simple form of the metric on
moduli space in the ${\cal N}$=2$^*$ solutions, will be of use in the investigation
and construction of dualities in the future. Our work has
hopefully also added to the understanding of the power of brane probing
as a tool in such investigations.

\vskip .2in
\noindent
{\bf Acknowledgements}\vskip .1in
\noindent
The authors are grateful to Clifford Johnson, Michela Petrini and Nick
Warner  for discussions and
comments on the manuscript.  
N.E is grateful for the support of a PPARC Advanced Fellowship,
J.B. for the support of a PPARC Studentship, and J.H.
for the support of the Department of Education, Isle of Man and
his parents.

\end{document}